\documentclass{aa}  

\usepackage{graphicx}
\usepackage{txfonts}
\begin{document}

   \title{Revisiting V1309 Sco 2008 outburst spectra}

   \subtitle{Observational evidence for theoretical modeling of stellar mergers}

   \author{Elena Mason
          \inst{1}
          \and
          Steven N. Shore \inst{2}
          }

   \institute{INAF-OATS,
              Via G.B. Tiepolo 11, 34143 Trieste, Italy\\
              \email{elena.mason@inaf.it}
         \and
             Dipartimento di Fisica "Enrico Fermi", Universit\`a di Pisa and INFN -- Sezione di Pisa, Largo B. Pontecorvo 3, 56127 Pisa, Italy\\
             \email{steven.neil.shore@unipi.it}
        }

   \date{Received September 15, 1996; accepted March 16, 1997}

% \abstract{}{}{}{}{} 
% 5 {} token are mandatory
 
  \abstract
  % context heading (optional)
  % {} leave it empty if necessary  
   {V1309 Sco is the only certain noncompact stellar merger, due to its indisputable preoutburst light curve matching that of a contact binary of almost equal mass stars. Therefore, anything that can be deduced from the existing observations serves as benchmark constraints for models.}
  % aims heading (mandatory)
   {We present some observational evidences to guide future hydrodynamical simulations and common envelope studies.  }
  % methods heading (mandatory)
   {Using archive spectra taken at high and mid spectral resolution during the V1309 Sco outburst and late decline, together with the inferential methods we developed to study nova ejecta through panchromatic high resolution spectroscopic follow ups, we constrain the physical state, structure, dynamics and geometry of the transient originated in the stellar merger.  }
  % results heading (mandatory)
   {We found that the emitted spectra arise from two distinct contributions: matter expelled during the 2008 outburst and circumbinary gas produced during historic mass loss episodes. These two components likely have orthogonal geometry with the 2008 mass loss displaying a dust-laden bipolar ejecta produced by a time limited rapidly accelerating wind and the circumbinary gas having a donut-like shape. A central source powers them both, having produced a fluorescent light pulse, but we cannot precisely determine the time it started or its spectral energy distribution. We can, however, place its upper energy cutoff at about 54 eV and the bulk of its emission at $<$20 eV. We also know that the central source turned off within months from the outburst and before the ejecta turned optically thin.}
  % conclusions heading (optional), leave it empty if necessary 
   {}


   \maketitle
%
%-------------------------------------------------------------------

\section{Introduction}

Discovered as a transient in 2008 (e.g. Nakano et al. 2008), V1309 Sco was first thought to be a microlensing event (Bensby, private communication). It was soon classified as a classical nova (Naito \& Fujii 2008). 
Monitored during its early decline through high resolution spectroscopy at the VLT+UVES, V1309 Sco was suggested to be a so called "red-nova" (Mason et al. 2010) and eventually shown to be a non-compact stellar merger (Tylenda et al. 2011). Here, we re-visit the UVES spectra published in Mason et al. (2010), together with the late decline sequence of VLT+XShooter spectra published in Kaminski et al. (2015). 

Over the past several years we have studied panchromatic high resolution spectra of classical novae unveiling the dynamic of their ejecta. Our intent is to use that experience to discern the dynamics and the physics of the stellar merger, and to produce empirical constraints to guide numerical simulations.   
%--------------------------------------------------------------------
\section{The ESO archive data}
The V1309 Sco spectra analyzed herein are in the European Southern Observatory (ESO) archives. These correspond to six UltraViolet Echelle Spectrograph (UVES) epochs taken with the DIC1(346+580) and DIC2(437+860) settings which, together, cover the wavelength range $\sim$3000-10300\ \AA \ in each epoch; plus 3 XShooter epochs covering the wavelength range $\sim$3000 \AA \ - 2.4 $\mu$m each. 
The UVES data were obtained during regular service mode operations, where downloaded from the archive\footnote{http://archive.eso.org/eso/eso\_archive\_main.html} and reduced on a local computer using the {\it esorex} pipeline (version 6.1.3) with {\it gasgano} interface (version 2.4.8). The data were flux calibrated using the "master response" function produced by the {\it quality control group}, rather than the "instrument response" function that can be generated with the spectrophotometric standard star observation of the night (if taken), because experience showed that the spectra are better calibrated in the whole spectral range when the master response is used. 
The XShooter data were, instead, taken during commissioning and science verification\footnote{https://www.eso.org/sci/facilities/paranal/instruments/xshooter/news.html and links therein (bottom page).% https://www.eso.org/sci/activities/vltsv/xshootersv.html
}. The reduced spectra are those already published in Kaminski et al. (2015) that provide the details of the data reduction process. Here we add that the extraction of the third XShooter spectrum was done using a reduced extraction window, smaller than the spectrum spatial full width half maximum (FWHM), to avoid the part contaminated by a nearby star (the seeing allowed such a crude deblending). 
Adopting the time of the transient's discovery, 2008 Sep. 2.46 UT, as $t_0$, the UVES spectra sample the first decline phase, the plateau and the subsequent steep decline phase displayed by the V band light curve (Figure 1 in Mason et al. 2010). The XShooter spectra, taken about one year later, monitor the late decline (e.g. Figure 2 in Kaminski et al. 2015). The observing date and the age of the transient at each epoch are in Table~\ref{table:log}. 
%-------------------------------------------------------------
%
\begin{table}
\caption{The date of the UVES and XShooter observations, together with the age of the transient at each epoch (considering 2008 Sep. 2.46, i.e. MJD 54711.46, the time of discovery, the $t_0$ epoch).}            % title of Table
\label{table:log}      % is used to refer this table in the text
\centering      % used for centering table
\scriptsize
\begin{tabular}{c c r c}        % centered columns (4 columns)
\hline\hline                 % inserts double horizontal lines
Obs. Date (UT) & Epoch \# & Age (d) & Instrument \\    % table heading 
\hline      % inserts single horizontal line
2008-09-13 & 1 & 10.6 & UVES \\ 
2008-09-18 & 2 & 15.6 & UVES \\
2008-09-20 & 3 & 18.5 & UVES \\
2008-09-28 & 4 & 25.6 & UVES \\
2008-10-08 & 5 & 35.6 & UVES \\
2008-10-20 & 6 & 47.6 & UVES \\
2009-05-04 & 7 & 243.7 & XShooter \\
2009-08-13 & 8 & 344.7 & XShooter \\
2009-09-29 & 9 & 391.6 & XShooter \\
\hline                    %inserts single line
\end{tabular}
\end{table}
%
%--------------------------------------------------------------------
\section{Analysis of the UVES data set}

%One column figure
%----------------------------------------------------------------- 
   \begin{figure}
   \centering
   \includegraphics[width=8.5cm]{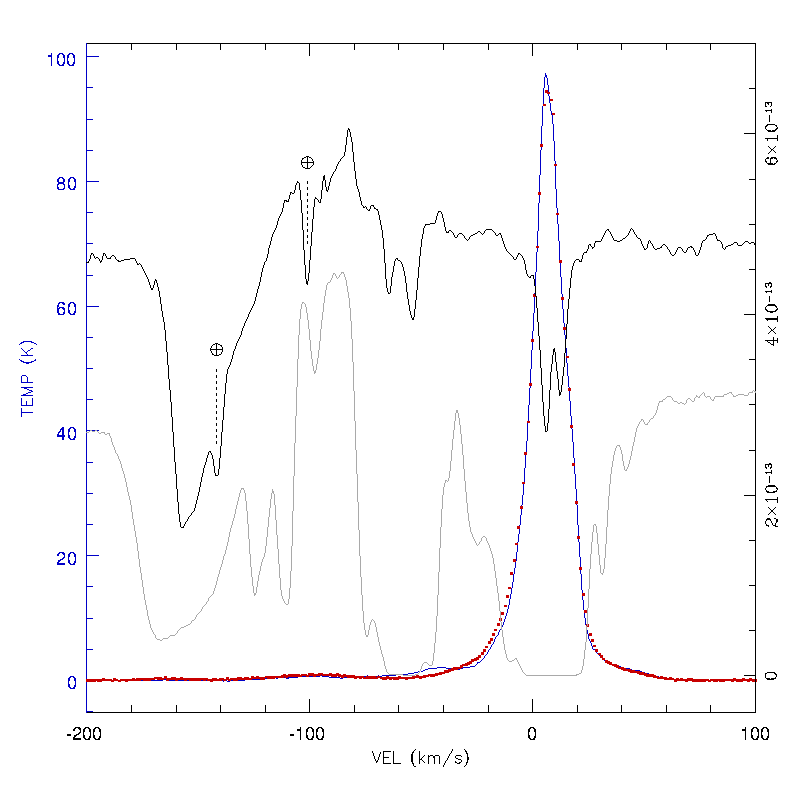}
      \caption{GASS/LAB H~I 21 cm temperature profile (solid blue line and red dots for the GASS and LAB measurements, respectively), compared with the K~I $\lambda$7699 (black line) and Na~I~D (gray line) interstellar absorptions. The units of the right vertical axis are erg/cm$^2$/s/\AA. Telluric absorptions are marked with the $\oplus$ symbol. Velocities are in the Local Standard of Rest (LSR) reference frame. 
              }
         \label{fig:reddening}
   \end{figure}
%----------------------------------------------------------------
\subsection{Reddening and distance}
The reddening for V1309 Sco cannot be inferred from the Na~I or Ca~II interstellar absorption lines since these are saturated and severely blended with low velocity absorption structures from V1309 Sco itself. However, this is not the case for the K~I absorption doublet which is clean and traces the same interstellar gas as Na~I. Fig.\ref{fig:reddening} compares the Galactic-All-Sky-Survey/Leiden-Argentine-Bonn (GASS/LAB) 21 cm temperature profile (Kalberla et al. 2005)  with the absorption line profile from the K~I $\lambda$7698.96 line in the first UVES spectrum. The comparison suggests that the bulk of the neutral H mapped by GASS/LAB is  between V1309 Sco and the observer and that we can integrate the whole emission profile between $-40$ and $+$65 km/s for a reliable estimate of the reddening. We derive a H~I column density of $\sim$3.7$\times$10$^{21}$ cm$^{-2}$ and E(B-V)=0.65$\pm$0.01 mag. 

The K~I and Na~I profiles show evidence for an additional absorption component centered at v$_{LSR}\sim-$60 km/s  which does not have a corresponding feature in the GASS/LAB profile. 
We favor the interpretation that they are of interstellar origin  since we observe them in all the ground state transitions (e.g. Na~I, K~I, etc) but in none of the transitions from ionized atoms (e.g. Fe~II, Sc~II, etc)  that certainly originate in the transient's gas. The difference between the metal atomic absorptions and the H~I profile can be explained by the patchy line of sight toward the Galactic center and the coordinate offset between the V1309 Sco the H~I surveys pointings, that can exclude or include intervening clouds. The LAB nearest pointing is offset by $\geq$0.21$^o$, while the effective beamsize is 0.266$^o$.  %https://www.astro.uni-bonn.de/hisurvey/profile/index.php
Expanding the range for the integral of the GASS/LAB  profile to $-80$ to $+65$ km/s to include such additional absorptions, does not change the resulting H~I column density and reddening.  

Using different diagnostics, however, the reddening changes depending on whether the $-60$ km/s absorptions are taken into account. In particular, measuring the K~I equivalent width, EW, and adopting the Munari \& Zwitter (1997) empirical relation, we get E(B$-$V)$\simeq$0.60$\pm$0.02 mag from the average of the measurements from the six UVES spectra for the absorptions centered at $\sim0$ km/s alone, and E(B$-$V)$\simeq$0.84$\pm$0.04  when including the $\sim-60$ km/s absorptions.

Finally, we can obtain a reddening estimate using the EW of diffuse interstellar bands (DIBs). Although most of the strongest DIBs are either severely blended with absorption features from the transient or in the wavelength gaps of the UVES setup, we can measure the EW of the DIB at 6613.3 \AA \ and  6283.8 \AA, albeit with a lower confidence because of the broad profile and the imperfect telluric correction. Averaging the measurements from all six UVES spectra we get EW$\simeq$0.15$\pm$0.03 \AA \ for the 6613 \AA \ DIB and EW$\simeq$0.64$\pm$0.14 for the 6284 \AA \ DIB. They correspond to E(B-V) in the range 0.67-0.73 and 0.54-1.03, respectively, depending on the calibration (e.g. Jenniskens \& Desert 1994, Friedman et al. 2011). 
The two DIBs match, in velocity, the positions of the $\sim0$ km/s clouds in the LSR, possibly indicating a smaller column density for the cloud at $\sim-60$ km/s. 

Noting that these three methods are completely independent, we adopt 0.65$\leq$E(B-V)$\leq$0.84 mag and work with these two extremes, whenever the dereddening correction is needed to the analysis. 

The additional interstellar gas cloud at LSR velocity $\sim-60$ km/s, has an interesting implication for the distance of V1309 Sco.  
The V1309 Sco coordinates point toward the Galactic center ($l\sim360$ and $b\sim-3$ deg), at which the Dame et al. (2001) CO maps show the "3 kpc ring" molecular gas velocities compatible with the observed $-60$ km/s. This implies that the transient is behind the "3 kpc molecular ring" and therefore at about $\sim$ 5 kpc from the Sun. This is somewhat farther than the 3-3.5 kpc previously estimated (e.g. Tylenda et al. 2011). 

%-------------------------------------- Two column figure (place early!)
   \begin{figure*}
   \centering
   \includegraphics[width=8cm]{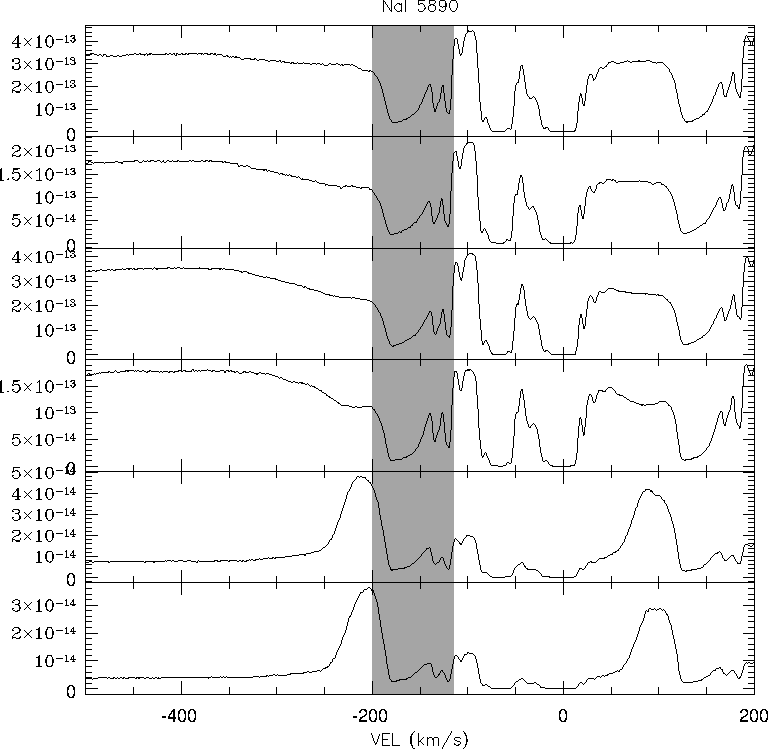}
   \includegraphics[width=8cm]{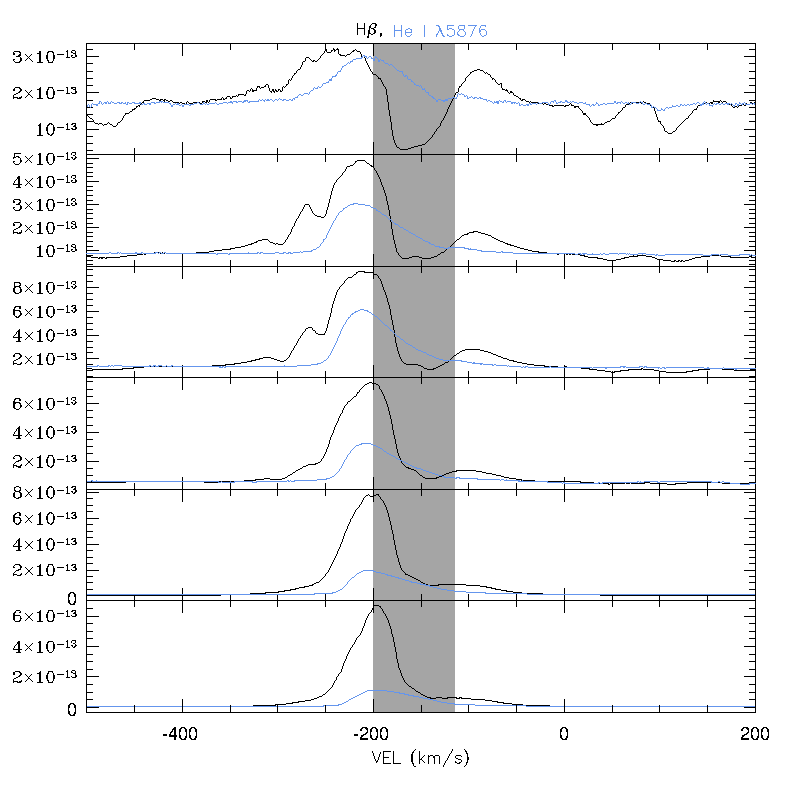}
   \includegraphics[width=8cm]{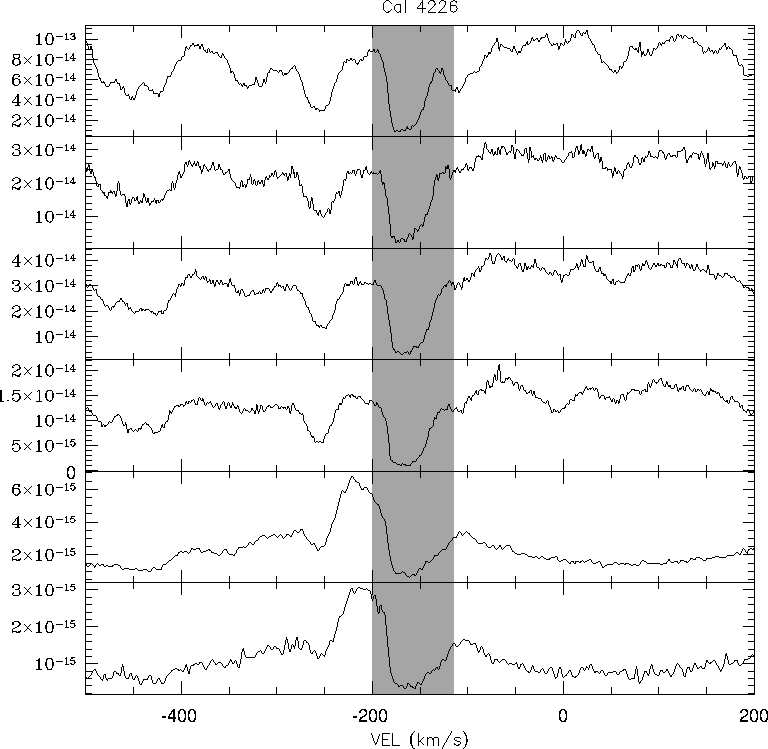}
   \includegraphics[width=8cm]{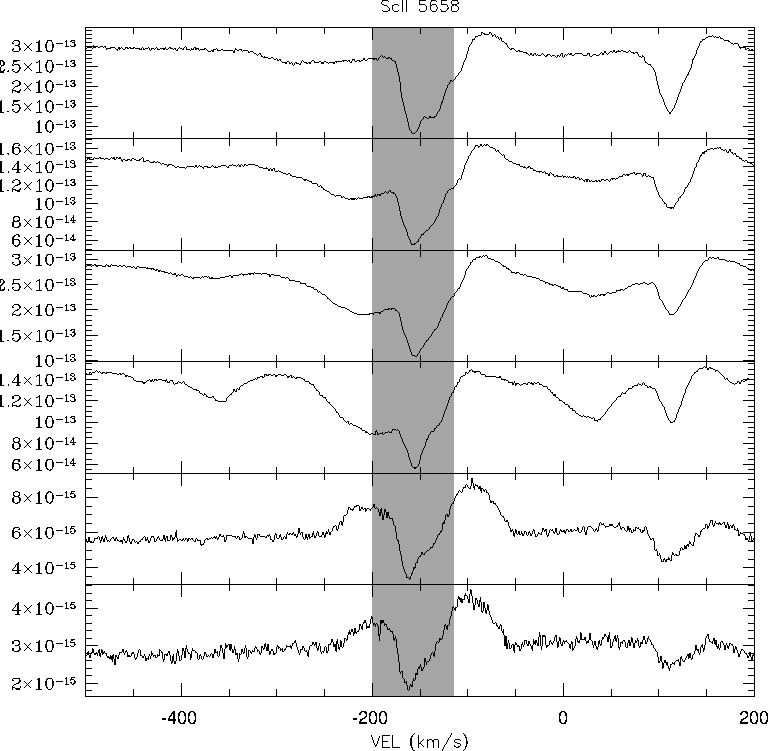}
   \includegraphics[width=8cm]{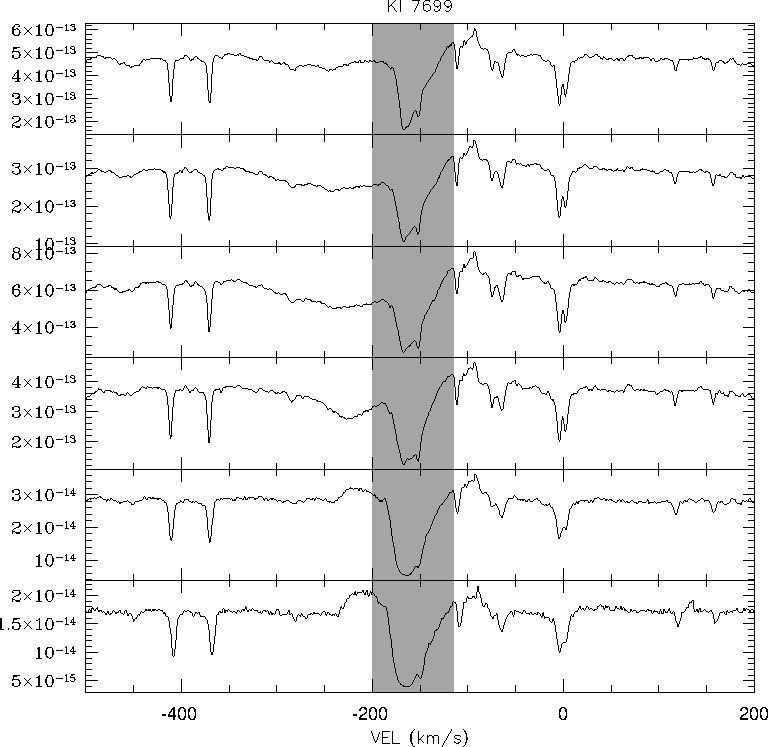}
   \includegraphics[width=8cm]{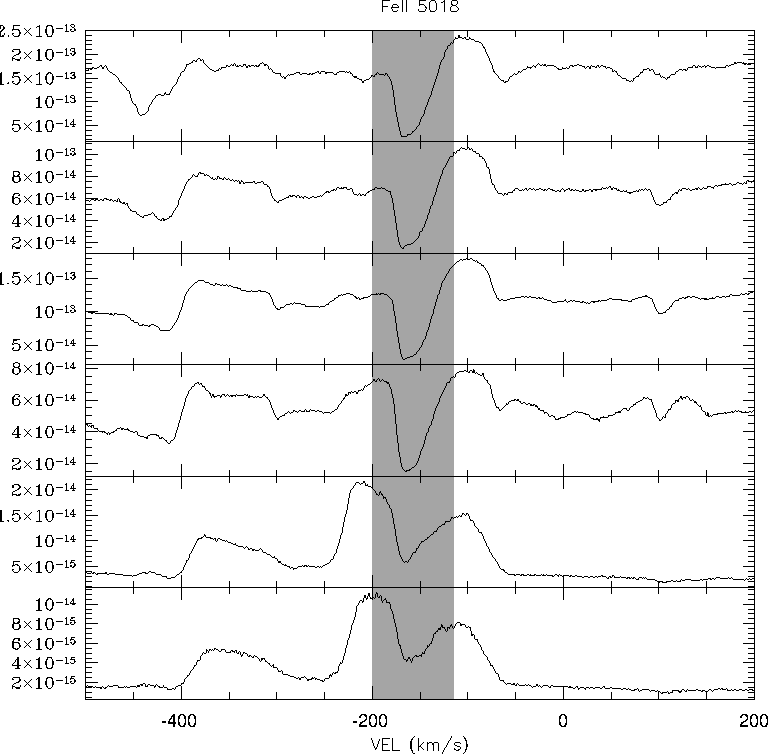}
   \caption{Examples of line profiles and their evolution across the UVES spectra. Each subpanel is for a different transition (specified in the subpanel title); early epochs are on top, and later ones in the bottom of each subpanel. The vertical shaded area marks the velocity range (-200,-115) km/s where the persistent absorption is located. Velocities are in the heliocentric rest frame. Y-axis units are in erg/cm$^2$/s/\AA.}
              \label{fig:persistent_abs}%
    \end{figure*}
%------------------------------------------------
\begin{figure*}
    \centering
    \includegraphics[width=8cm]{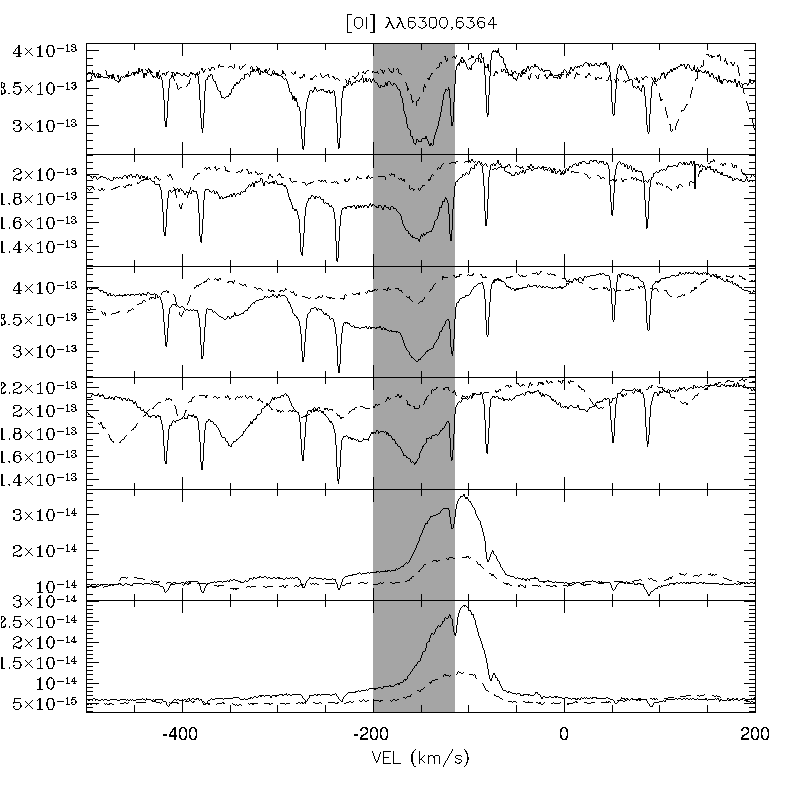}
    \includegraphics[width=8cm]{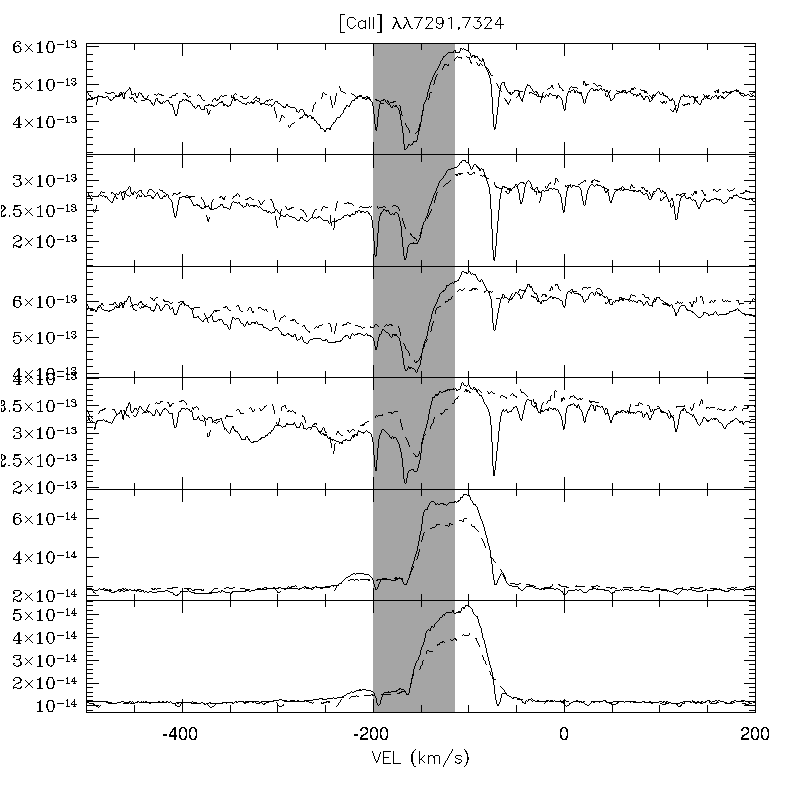}
    \caption{Line profiles of the forbidden doublets [O~I] and [Ca~II] (the dashed line is for the weakest of the transitions in the doublet). Times increase from top to bottom within each subpanel; while the shaded area marks the velocity range (-200,-115) km/s where persistent absorptions are observed. Velocities are in the heliocentric rest frame. Y-axis units are in erg/cm$^2$/s/\AA.}
    \label{fig:forbidden_abs}
\end{figure*}
%------------------------------------------------

\subsection{The line profiles, their evolution and the inferred structure}\label{sec:lineprofile}
Figures~\ref{fig:persistent_abs} and \ref{fig:forbidden_abs} show a selection of line profiles and their evolution across the six UVES epochs. 
We distinguish two different portions of each line in velocity space: 1) the range  ($-$360,$-$200) km/s, which we call the "high velocity component" (HVC); and 2) the range (--200,--50) km/s which we call the "low velocity component" (LVC) and which displays what looks like a persistent absorption in the velocity range (--200,--115) km/s (the gray band in Figs. \ref{fig:persistent_abs} and \ref{fig:forbidden_abs}). These  velocity limits are approximate, since they may vary slightly with the transition. The two ranges manifest different evolution. Their line ratios are also substantially different, suggesting different opacity and gas conditions. 

The HVC is constantly evolving in our sequence of spectra. It appears as a weak, shallow and  asymmetric absorption that progressively increases in strength from epoch 1 to 4  and eventually turns into a strong emission profile (from epoch 5 on), which peaks at $\sim-$210 km/s and has an extended blue wing. The strength of the emission relative to the continuum, also increases with time.  
This evolution is particularly evident in relatively isolated transitions (i.e. free from confounding blends) that are either resonance lines (e.g. Na~I and K~I panels in Fig.~\ref{fig:persistent_abs}) or have metastable lower energy levels (e.g. Sc~II panel in Fig.~\ref{fig:persistent_abs}). Recombination lines, such as the early H Balmer and He~I transitions, are always in emission (or emission dominated), with the line strength steadily increasing from epoch 1 to 6 and later (XShooter spectra), relative to the continuum. 
Because of its velocity, we associate the HVC with gas expelled from the system during the outburst. Its evolution is also consistent with gas whose physical conditions of opacity, ionization, and possibly density, change because of expansion. The shallow absorption initially increases in strength because of the of a propagating recombination wave (Shore et al. 2011, Shore 2012); it then turns into emission once the optical depth has substantially decreased. Similarly, the strength of the H and He~I high velocity components grows constantly as self absorption effects decrease. 

%------------------------------------------------
\begin{figure*}
    \centering
    \includegraphics[width=14cm,angle=270]{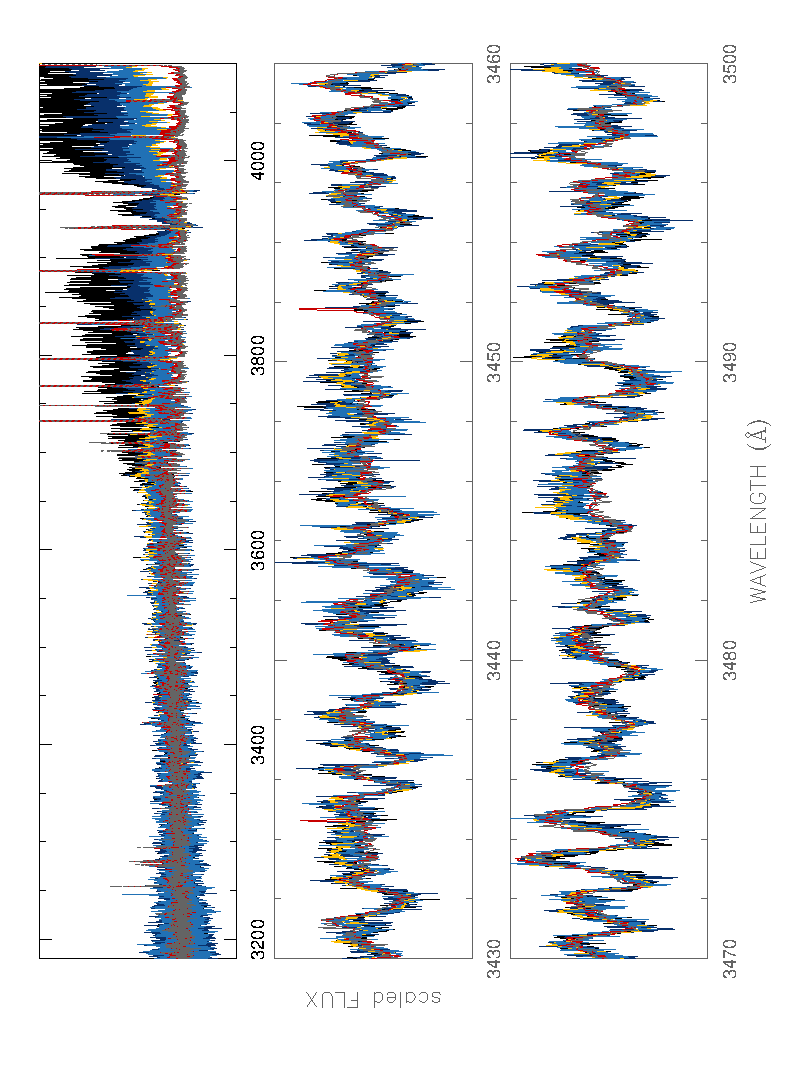}
    \caption{Selected ranges in the U-band spectrum showing the "curtain". This is mostly produced by ground transitions from neutral and singly ionized metals and completely absorbing the continuum blue-ward of the Balmer jump. The top panel shows the whole wavelength range 3200-4100 \AA; the middle and the bottom panels show zoomed in portions of the curtain. Black, blue, yellow, light-blue, red and dotted-gray are for epoch 1, 2, 3, 4, 5 and 6, respectively.}
    \label{fig:curtain}
\end{figure*}
%------------------------------------------------
\begin{figure}
    \centering
    \includegraphics[width=8.5cm]{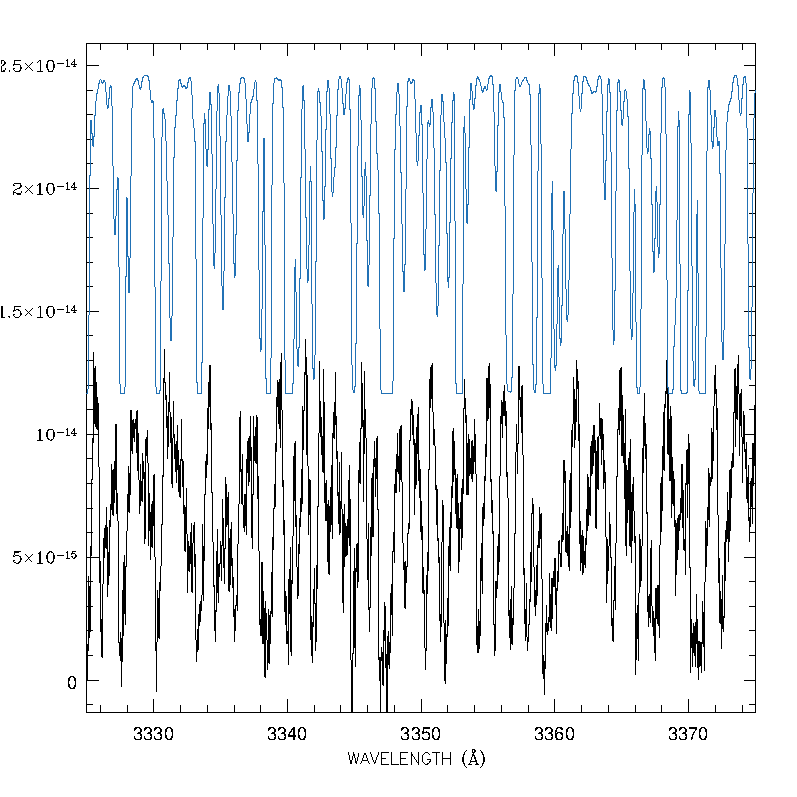}
    \caption{Comparison between the observations (black line, epoch 1 spectrum, smoothed with a running box of 5) and the curtain simulation produced as described in the text (light blue line). Units of the y axis are erg/s/cm$^2$/\AA. The simulation has been arbitrarily scaled. }
    \label{fig:curtainsim}
\end{figure}
%------------------------------------------------
The LVC consists of a ubiquitous absorption component and a relatively weak emission that together resemble a P Cyg profile (where blending with other transitions is not affecting significantly the whole profile). On close inspection, the absorption component is persistent and almost invariant in width, profile and, to a less extent, depth, especially in the resonance transitions in the first four epochs. Moreover, the 3200-3700 \AA \ wavelength range displays a plethora of neutral and singly ionized metal ground and low excitation energy transitions that form a curtain absorbing the continuum radiation during all six UVES epochs (see Fig.\ref{fig:curtain}). Those absorptions do not change in time, what changes is the continuum. This is why we dubbed them "persistent absorptions".  They are also observed in the [O~I] and [Ca~II] doublets and this has interesting physical implications that we discuss in Sect.~\ref{sect:Ne}.
To test the effective presence of a curtain we performed a simple simulation to compare with our data: we selected transitions from  the neutral and singly ionized metals Fe, Ti, and Cr, using the Kurucz's line list (Kurucz \& Bell 1995), and computed their opacity assuming local thermodynamic equilibrium (LTE) conditions, an excitation temperature T$_{ex}$=10$^4$ K, and solar abundances. No thermal broadening was assumed and a uniform dynamical broadening of 8 km/s was applied to all, in agreement with the observed lines width. Each line was weighted only by its energy level and degeneracy, and the elemental abundance. By including only 3 elements and assuming LTE conditions, when in fact the we are probably observing the relics of a wind (see also  Sect.~\ref{sect:end}), we are far from model fitting the data. Instead, we demonstrate that the LVC is consistent with a curtain, i.e. passive gas absorbing incident radiation. In particular, Fig.~\ref{fig:curtainsim} shows that even such a crude simulation is capable of reproducing the observed pattern of absorptions in the data. 

\begin{figure}
   \centering
   \includegraphics[width=8.5cm]{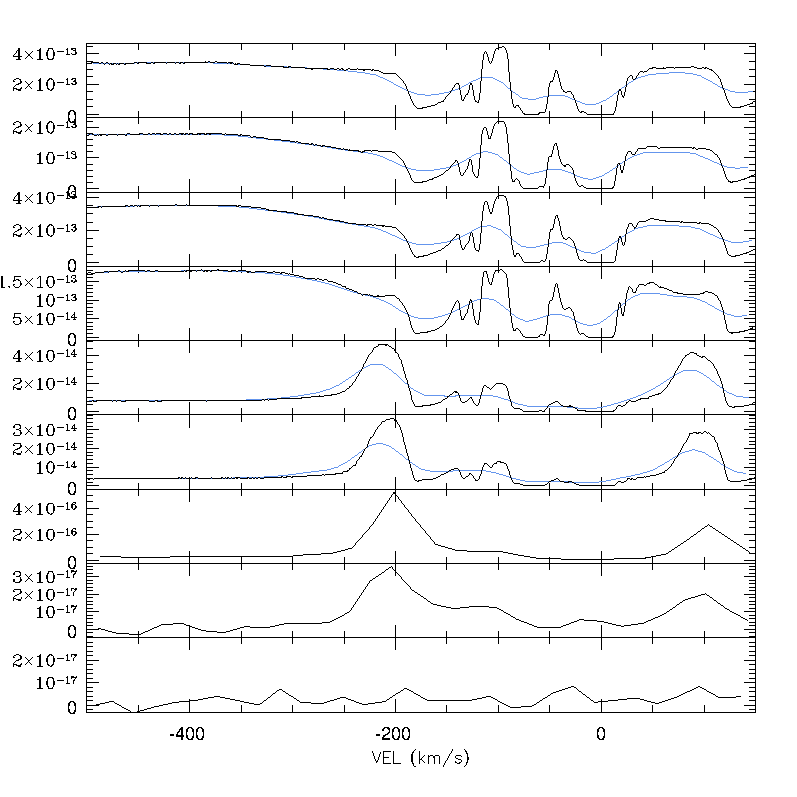}
  \caption{Na~I~D emission lines evolution across time (time increase from top to bottom). The Black color is for the original spectra (UVES or XShooter), while the light blue is for the UVES spectra degraded to the XShooter resolution. Comparing epochs 5 and 6 (UVES) with epochs 7 and 8, it is evident that what appear as a "red wing plateau" in low resolution spectra, could well be a signature of the non-resolved persistent absorption. Velocities are in the heliocentric rest frame. Y-axis units are erg/cm$^2$/s/\AA.
              }
         \label{fig:NaIpersistentABS}
   \end{figure}

The curtain and its persistence suggest that the absorbing gas is not part of the 2008 mass loss but, rather, circumbinary material. 
This picture is also supported by the subsequent evolution of the whole LVC. Although the detailed evolution of the profiles appears different for different ions, the emission component generally increases over the continuum while the absorption decreases and appears either in front of the continuum or in front of the emission components (including the HVC), or both. This evolution is consistent with a change in the emissivity. There are no signatures of dynamical effects such as expanding gas, whose change in opacity would produce recombination waves, or shocking gas clouds, whose interaction would cause substantial line profile changes. Hence, we propose that the gas responsible for the LVC is material produced in past mass loss episodes, and that the increase in emissivity in this material results from a light pulse (the 2008 outburst) and its time delayed fluorescence and diffusion across the medium. The circumbinary gas is ionization bounded, at least in the time interval spanned by the six UVES epochs (as shown by the persistent absorptions), and probably beyond. Degrading the UVES spectra to the XShooter resolution and inspecting the Na~I~D line profile, it appears that partially neutral circumbinary material is also present at the time of the first two XShooter observations (see Fig.~\ref{fig:NaIpersistentABS}).  
This further indicates that such a circumbinary gas does not interact dynamically with the low velocity components at least until day 345. Moreover, since the XShooter epoch 9 spectrum shows no new transitions and is a fainter replica of the epoch 7 and 8 spectra (see Section~\ref{sect:xsh}), we conclude that there is no dynamical interaction between the LVC and the HVC until at least day 392. 

%One column figure
%----------------------------------------------------------------- 
%----------------------------------------------------------------

%One column figure
%----------------------------------------------------------------- 
   \begin{figure}
   \centering
   \includegraphics[width=8.5cm]{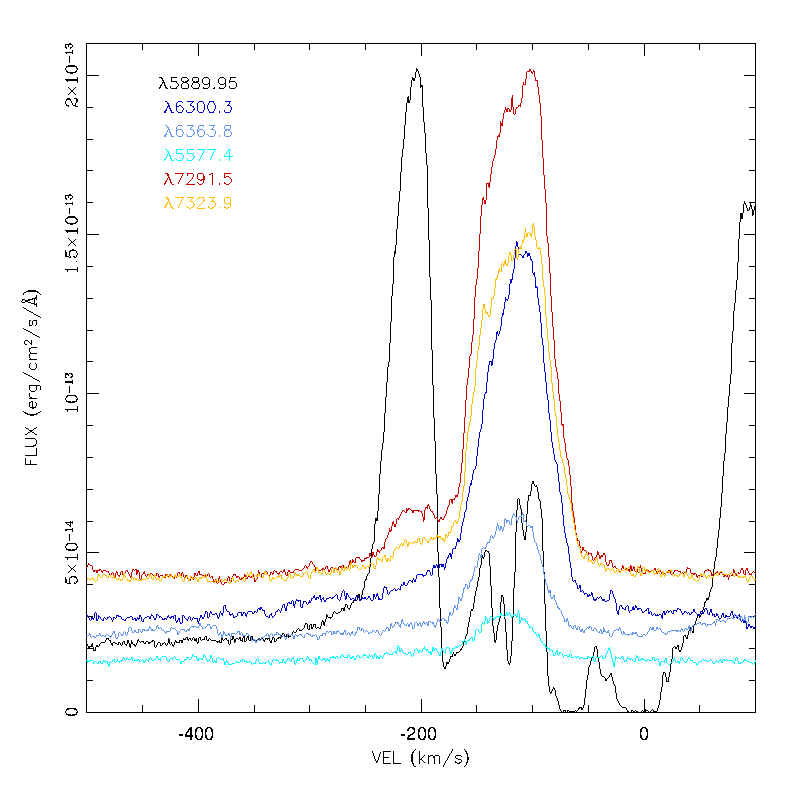}
  \caption{Comparison between the UVES epoch 6 Na~I~D and the forbidden transitions ([O~I] and [Ca~II]) profiles. The color code for the identification of each line is in the figure panel itself. The spectrum in the figure has been corrected for reddening and telluric absorptions. Velocities are in the heliocentric rest frame.
              }
         \label{fig:barycenter}
   \end{figure}
%----------------------------------------------------------------
A final, remarkable aspect of the observed line profiles is that the HVC completely lacks emission with positive velocities or better symmetric with respect to the systemic velocity (which we estimate to be $\sim-$120 km/s in Section~\ref{sect:gammavel}). Since we regard it unlikely that there was an asymmetric mass loss from just one stellar hemisphere, we suggest that the receding component of the outflow is obscured or rather self-obscured by dust within the ejecta.  Shore et al. (2018 and references therein) showed how the dust can suppress the receding part of an emission line profile in an axis-symmetric ejecta inclined to the line of sight. In this case, in order to hide the receding portion of the emission line, the dust must be well mixed in the ejecta and have a large opacity. %(of at least 10 mag?). 
Using the He~I $\lambda$5876 emission line in the epoch 5 spectrum, since it is HVC dominated (i.e. unbiased by the LVC), and the intensity ratio between the line peak and the point in the red wing which is symmetric around the peak with respect the systemic velocity, we find almost 4 mag (3.8 mag, precisely, and $\tau_d$=3.5) of dust extinction across the line. Using the He~I $\lambda$7065 emission to perform the same estimate we find 3 mag of dust extinction (or $\tau_d$=2.7) between the high velocity peak emission and its corresponding symmetric point with respect to the systemic velocity. Assuming silicate dust, as suggested by the 2010 observations of Nicholls et al. (2013), our observed $\tau_d$ ratio is consistent with the grains being $\gtrsim$0.1~$\mu$m already 35 days after discovery (using Draine \& Lee, 1984, absorption efficiency plot in their Figure 5). Repeating the exercise with the epoch 6 spectrum, we obtain $\tau_d(5875)$=3.6 and $\tau_d(7065)$=3.2 for a ratio of 1.13 which may indicate grain growth to sizes of order of $\sim$0.3~$\mu$m. We could not repeat the measurements and calculation in the first four spectra, since, although the He~I lines are present, the $\lambda$7065 red wing is biased by partial blending with adjacent emission lines, while the He~I $\lambda$5876 shows also the contribution of the LVC emission in its red wing. 

The global picture, therefore, is that the HVC is mass lost during the 2008 outburst, and the LVC is surrounding circumbinary material expelled sometime in the past. Their exact geometry is not constrained at this stage. However, there are two possibilities: either the two structures are very distant to each other so that the HVC does not reach the low velocity gas within the time of the spectroscopic observations\footnote{For an expansion velocity of the HVC of $\sim$400 km/s (see Section~\ref{sect:end}) the very lower limit for the LVC's distance is 400$\times$392$\times$86400 km $\simeq$90 AU}; or they have orthogonal geometries so that they cannot interact regardless of their time of expulsion and relative distances. We will return to this in Section~\ref{sect:end}.  

\subsection{The systemic velocity}\label{sect:gammavel}
Having established the circumbinary nature of the LVC, some quantitative information can be derived if it has a minimal axisymmetric distribution around the central source (e.g. a spherical shell, an equatorial disk, etc.). 
Indeed, for optically thin lines, such as [O~I]$\lambda\lambda$6300,6364 and [Ca~II]$\lambda\lambda$7291,7323 that are fully in emission in epoch 5 and 6, the whole emitting volume is visible. Thus, the line centroids match the volume geometrical center where the merger is sited. Fig.~\ref{fig:barycenter} compares the profiles of the [O~I], [Ca~II] and Na~I transitions in the epoch 6 spectrum. It shows that the blue wing of the emission, i.e. the outermost layer of the low velocity gas, might still be unexcited (undetected), implying a bias toward lower values if the line centroid is used. To overcome such a bias we measured the [O~I]$\lambda$5577 line, that is not a ground state transition and is free from absorption components, and compared its flux centroid with that of the peak of the other forbidden transitions as well as the bisector between the blue wing of the early epoch absorption and the red wing of the late epoch emission. 
The resulting radial velocity of the system, $\gamma$, is between $-$120 and $-$125 km/s. Hence, the projected terminal velocity of -180 km/s observed in the Na~I~D persistent absorption reduces to $\sim-$60 km/s.
This value of $\gamma$ revises the $-$80 km/s estimated by Mason et al. (2010),  and the  $-$175 km/s reported by Kaminski et al. (2015), who, however, concluded that the derived velocity was unlikely representative of the system radial velocity. Our $\sim-$120 km/s heliocentric systemic velocity is in good agreement with the v$_{LSR}=-$100 km/s, v$_{helio}\sim-$110 km/s, measured by Kaminski et al. (2018). 

\subsection{The  column density of the low velocity component}\label{sect:Ne}
As mentioned in Section~\ref{sec:lineprofile}, the low velocity persistent absorption is also evident in the forbidden transitions [O~I]$\lambda\lambda$6300,6364, [Ca I]$\lambda$6527, and [Ca~II]$\lambda\lambda$7291,7323. The presence of an absorption component in forbidden transitions indicates a very large column density of the intervening gas. 
After removing the telluric absorptions around the [Ca~II] and [O~I] lines using {\it molecfit} (Smette et al 2015, Kausch et al. 2015) and {\it IRAF/telluric}, we followed Savage \& Sembach (1991) to compute the O~I, Ca~I and Ca~II column density. Estimated column density for each absorption and each epoch are listed in Table~\ref{table:N}. Uncertainties are of the order of $few \times10^{14}$ and $few \times10^{16}$ for the [Ca~I] and [Ca~II] transitions, respectively, and $few \times10^{19-20}$ for the [O~I]. The larger uncertainties in the [O~I] are explained by the ambiguity on the local continuum level position (given the weak emission component) and the suspected presence of blending features (e.g. [Ni~I], see Allende Prieto et al. 2001). Using the equivalent width relation for optically thin gas (e.g. Spitzer 1968), we obtained similar values for the Ca$^{0,+}$ and the O$^0$ column density.  
Adopting Asplund et al. (2021) solar abundances for Ca and assuming that calcium is either neutral or singly ionized, we derive N$_H\sim$2.5$\times10^{23}$ cm$^{-2}$. Similarly, we derive N$_H\sim$5.3$\times10^{23}$ cm$^{-2}$ from the average of the N$_O$ measurements. A factor of 2 uncertainty is compatible with the method and the error in the measurements. 

%-------------------------------------------------------------
%                                             Simple A&A Table
%-------------------------------------------------------------
%
\begin{table}
\caption{The Ca, Ca$^+$ and O column density in cm$^{-2}$.}             % title of Table
\label{table:N}      % is used to refer this table in the text
\centering      % used for centering table
\scriptsize
\begin{tabular}{l c c c c c}        % centered columns (4 columns)
\hline\hline                 % inserts double horizontal lines
Epoch/Age & [Ca~II]$\lambda$7323 & [Ca~II]$\lambda$7291 & [O~I]$\lambda$6364 & [O~I]$\lambda$6300 & [Ca~I]$\lambda$6573 \\    % table heading 
\hline      % inserts single horizontal line
1/10 &  5.0e17 & 5.0e17 & 3.4e20 & 3.8e20 & 8.0e13 \\  % inserting body of the table
2/15 &  5.5e17 & 4.0e17 & 2.0e20 & 2.2e20 & 1.3e14 \\
3/17 &  5.5e17 & 3.7e17 & 1.9e20 & 2.3e20 & 1.4e14 \\
4/25 &  5.5e17 & 5.0e17 & 1.6e20 & 3.4e20 & 2.4e14 \\
\hline                    %inserts single line
\end{tabular}
\end{table}
%--------------------------------------------------------
\subsection{The light pulse energy distribution}\label{sect:pulse}

The low velocity gas absorbs in the neutral and singly ionized metals such as Fe, Sc, Ti, V, Cr, Sr, Ba and most importantly Ca, the neutral (alkali metals) Na and K, and neutral (non metals) O and H. 
All these transitions are either ground transitions or produced by pumping mechanisms, or fluorescence. 

The H Balmer line absorptions indicate a very opaque Ly$\alpha$ capable of substantially populating the $n=2$ energy level. The emission part, instead, traces recombination in addition to decay from the excited levels, but such an emission is minimal in the LVC. 
Fe, Sc, etc, because of their low ionization potential energies are mainly singly ionized. They belong to the so called iron-curtain elements whose resonance transitions in the UV range pump optical emissions with metastable low energy levels (Shore \& Aufdenberg 1993). The emission component of these transitions is relatively strong over the continuum just because of such a pumping mechanism. The absorption is also quite strong because of the overpopulation of the metastable low energy levels. 
As already noted by Mason et al. (2010), a pumping mechanism also occurs in the case of the Ca~II transitions and explains the relatively strong emission component of the [Ca~II] doublet compared to [O~I]. The Ca~II~H\&K transitions are extremely opaque (saturated) and overpopulate their upper level to the point that a detectable number of de-excitation occurs for the alternative channel of the NIR Ca~II triplet and the resonance [Ca~II] doublet.   

In contrast, the evolution of the oxygen transitions illustrates how the fluorescence strengthens with time. The [O~I] doublet upper level is populated either by collisions or by the emission of the intercombination line O~I]~$\lambda$1641, pumped directly by the ground transition $\lambda$1302 or indirectly by the fluorescent match of the Ly$_\beta$ and the O~I~$\lambda$1025 lines (see Shore \& Wahlgren 2010). The minimal emission component of the [O~I] doublet in the first four UVES epochs, together with the lack of the [O~I]$\lambda$5577 line\footnote{The [O~I]$\lambda$5577 emission appears in the epoch 5 UVES spectrum and persists across the XShooter spectra.} and the relative weakness of the emission component of the triplets at $\sim$8446 \AA \ (fluorescent) and $\sim$7774 \AA \ (recombination), suggest that fluorescence and recombination are only weakly occurring. They become important in the epoch 5 and 6 spectra, once the incoming radiation has progressed through and affected a substantial fraction of the circumbinary material. 
That some recombination is, however, occurring since at least epoch 1, is shown by the very weak LVC emission at the He~I $\lambda$5876 in the first four UVES epochs\footnote{The line strengthens until epoch 3 and declines afterward. In epoch 5 and 6 if present, the LVC He~I emission is at the level of or below the red wing of the HVC emission.}. Its weakness, together with the lack of other LVC He~I emissions indicate that very little He has been ionized and is recombining. This implies either that the incident He-ionizing radiation is minimal, perhaps diluted by a distance effect, or that the hydrogen is extremely optically thick and absorbs the vast majority of the photons with high enough energy to ionize helium. 
Either way, the set of observed lines and their profile suggest that the radiation reaching the low velocity gas has a high energy tail up to $\gtrsim$24 eV, but the bulk of its distribution peaks well below 20 eV. This is further supported by the lack of C~I and N~I LVC emissions that, if present would indicate recombination (IP$\sim$11 and 15 eV for the C and N, respectively) and of [N~II] absorption that suggest there is no ionized nitrogen (the [N~II] $\log(gf)$ is close to that of [O~I], while the N abundance is much greater than that of Ca), although for the nitrogen also apply the same reasoning than for the helium, i.e. it is little to no ionized because of the extreme opacity of the hydrogen. 

We also note that the He~I $\lambda$5876 reaches maximum strength in the epoch 3 spectrum and that in epoch 4 the line flux has already decreased by a factor $\exp(-1)$ indicating a recombination time of $\sim$ 8 days. This, using Hummer \& Storey (1998) He~I recombination coefficients, indicates electron densities of the order of n$_e\sim$7$\times$10$^7$ cm$^{-3}$. 

\begin{figure}
   \centering
   \includegraphics[width=8.5cm]{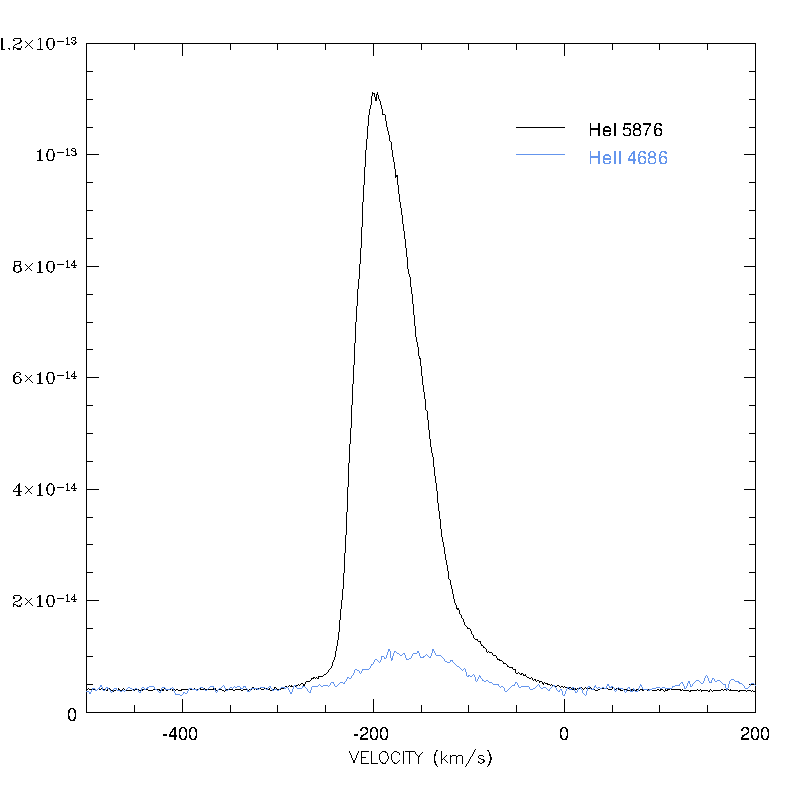}
\caption{Comparison of the He~I (black) and He~II (light blue) line profiles in the epoch 6 spectrum. The He~II line has been scaled by a factor 4. The vertical axis is flux in unit of erg/cm$^2$/s/\AA. Velocities are in the heliocentric rest frame. See text for details. 
              }
         \label{fig:2he}
   \end{figure}

The HVC is more ionized since it displays both He~I and He~II emissions. The He~I lines are already present in the epoch 1 spectrum and strengthen with time until epoch 3, declining afterward. They eventually disappear in the XShooter spectra sequence taken about one year after outburst. The He~II $\lambda$4686 emission, instead, is weak, but present, only in the epochs 5 and 6 spectra. This underlines once more the different conditions in the high velocity gas with respect to the LVC. The former is initially opaque with no emission except for the H Balmer series and the He~I lines. In particular, it is too opaque to produce He$^{+2}$. Only once the HVC gas has expanded sufficiently does the He~II emission line appear. Also, the velocity profile of the $\lambda$4686 does not  match that of the He~I lines but is weighted toward the low velocity range of the HVC (see Fig.\ref{fig:2he}). In contrast to an explosive ejecta where the higher ionization emissions appear (first) in the ejecta outskirts where the density is lower, the ionization structure here is similar to that of the Stromgren sphere and the density gradient, if any, is not as steep. Hence, the HVC is also irradiated by an underlying source and either receives more flux or see a "harder" source than the LVC gas. In the first case the low and the high velocity component are far apart, in the second case the low velocity component does not directly see the central source. We have no way to tell the two scenarios apart. 
%One column figure
%----------------------------------------------------------------- 
%----------------------------------------------------------------

The detection of the He~II $\lambda$4686 line increases the upper energy cutoff of the central source to $\sim$55 eV, in the extreme UV. Such high energies are sufficient to produce also  C$^{+3}$, N$^{+3}$ and O$^{+3}$ but we do not see C~III, N~III or O~III emission lines (e.g. $\lambda$4650, $\lambda$4640, and $\lambda$3047 and $\lambda$3759, respectively), as expected from recombination. Their nondetection could simply result from a low emission measure - after all the He~II~$\lambda$4686 is less than 2\% of the He~I~$\lambda$5876. This suggests that the central source spectral energy distribution extends into the EUV range, but peaks at $\sim$20 eV or less.  Either way, the disappearance of the He~I and He~II emission lines in the XShooter spectra suggests that the central ionizing source has turned off sometime between day +48 and day +244. 

%------------------------------------------------
\begin{figure*}
    \centering
    \includegraphics[width=16cm,angle=0]{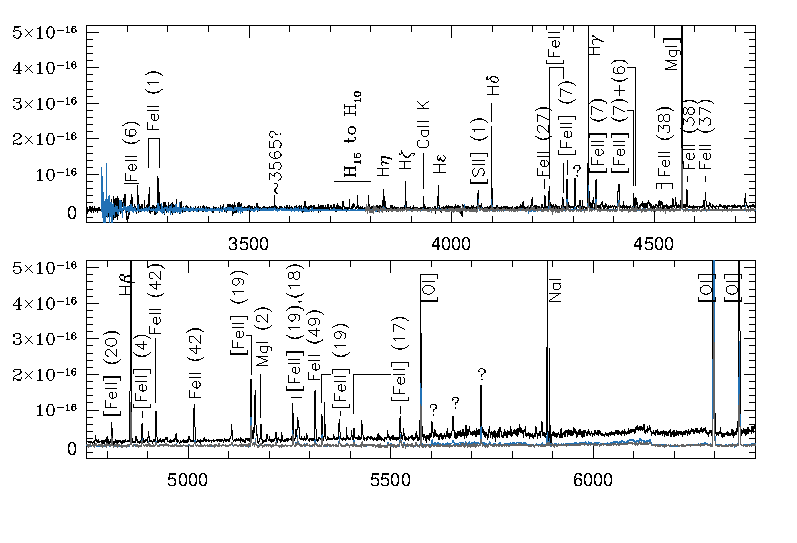}
    \caption{U to V band of the XShooter spectra. Black, light blue, and gray lines are for epoch 7, 8 and 9, respectively. Y-axis units are erg/cm$^2$/s/\AA; X axis units are \AA. The epoch 9 spectrum was cut below $\sim$3800 \AA, because no signal was evident there in the 2D frame.  The epoch 8 and 9 spectra are almost identical in the continuum (where present), with the epoch 9 spectrum having stronger molecular bands in emission. Bracketed numbers refer to the Moore's multiplet identification (Moore 1945).}
    \label{fig:curtain}
\end{figure*}
%------------------------------------------------
%--------------------------------------------------------------------
\section{Analysis of the XShooter data set}\label{sect:xsh}
The XShooter data set consists of three epochs about one year after the outburst, at days +244, +345 and +392. By this time the spectra appearance has changed dramatically. 
The continuum in the UVB arm (3000-5500 \AA) gradually fades below the detection threshold of the exposures. It is still detected in the epoch 7 spectrum but consistent with zero in the subsequent two epochs. Hence, we cannot verify whether the curtain produced by the low ionization potential metals that we observed in the UVES spectra is still present. In general, however, the lower spectral resolution of XShooter compared to UVES prevents firm conclusions about what, if any, narrow absorption that might still be present and absorbing against the continuum. Our analysis is, therefore, limited to the emission line spectrum. But, as discussed in Sect.~\ref{sec:lineprofile} and shown in Fig.~\ref{fig:NaIpersistentABS}, it is quite possible that the LVC is still present in the epoch 7 and 8 spectra. In epoch 9 the Na~I~D emission has almost disappeared, so the presence the low velocity persistent absorption cannot be determined.

The emission line spectrum is produced mainly by the high velocity component. It is difficult to identify all the emission lines with certainty since at any given wavelength there are a number of candidate low ionization potential metal identification (van Hoof 2018, NIST database), and we have already seen numerous blends in the higher resolution UVES spectra. However, we can safely say that the majority of the observed emission are Fe~II and [Fe~II] transitions. The Fe~II lines were already present in the UVES spectra with the high velocity emission first appearing in the epoch 4 spectrum, reaching the observed maximum 10 days after and subsequently gradually fading or disappearing. The [Fe~II] emissions, absent in the UVES spectra, are observed in the epoch 7 XShooter spectrum when they have maximum strength. They slightly weaken in the subsequent epochs.  All the [Fe~II] emission are ground or very low lying level transitions (the energy of the lower level is in the range 0-0.23 eV), with an energy jump of  $<$3 eV; while the Fe~II lines have upper energy levels $<$6 eV and, de-exciting, feed the upper levels of the forbidden transitions. That is, the [Fe~II] lines arise both from collisional excitation and from cascade down from the low levels of the Fe~II transitions, and, therefore, seems consistent with just the dilution of the high velocity component in the circumbinary space.

Other forbidden transitions that appear in the HVC and strengthen in the XShooter epochs are Mg~I] $\lambda$4571, [O~I], [Ca II], and [S~II]. The [S II] is particularly interesting because the sulfur ionization potential ($\sim$10 eV) is lower than for the hydrogen and oxygen, but much higher than for iron and related metallic species. Hence, it sets a stringent limit on the hardness of the ionizing source since we do not see any forbidden [O~II].
Of the two [S~II] doublets $\lambda\lambda$4068,4076 and $\lambda\lambda$6716,6730 only $\lambda$4068 is fairly strong and with a good S/N; the others are weak and obvious only in the epoch 7 and 8 spectra. Measurements of the doublet at $\sim$6700 \AA \ are compromised by blending with a molecular band. Nonetheless, using the five level models from Keenan et al. (1996), our line ratios $\log(F(4068)/F(6716+6730))$ $\log(F(4076)/F(6716+6730))$ are beyond their density range for any adopted reddening, suggesting electron densities in excess of 10$^5$ cm$^{-3}$. The large F(4068)/F(4076) ratio, $\sim$2.5 to 4, depending on the epoch, suggests densities close or above the critical value ($n_{cri}\sim$2-3$\times$10$^{6}$ cm$^{-3}$)\footnote{The expected ratio in the low and high density limit is 2 and $\sim$5, respectively}. The fact that the ratio increases between epoch 7 (2.5) and 8 (4), suggests in addition that there is a density gradient in the high velocity component, with the emissivity of lowest density portion dropping in favor of that of the higher density part, although the mid-resolution of XShooter does not suffice to identify sub-components within the line. The oxygen F(5577)/F(6300) measured line ratio ($\sim$0.2), compared with theoretical multilevel balance predictions assuming that the ionization equilibrium is dominated by electron collisions and charge exchange reactions and not an external radiation field (e.g. Hamann 1994), also indicates $n_e\gtrapprox$10$^6$ cm$^{-3}$. Finally, solving the equation of the detailed balance in the three level approximation for the [O~I] $\lambda\lambda$6300,6364, $\lambda$5577, $\lambda$2972 in the high density limit, we also obtain densities in the range 2$\times$10$^6\leq n_e\leq$4$\times$10$^6$ cm$^{-3}$, with an obvious addition by the low velocity component in the red wing of the $\lambda$5577 line from epoch 8 on.  The estimated  densities increase to 4.6$\times$10$^6$ cm$^{-3}$ adopting E(B-V)=0.84 mag. Again the apparent increase of the density with time indicate the presence of a density gradient. 

%--------------------------------------------------------------------
\section{Discussion and conclusions}\label{sect:end}
%\subsubsection{Like a light echo}
%\subsection{The facts}
In the spirit of Sherlock Holmes {\it "We balance probabilities and choose the most likely. It is the scientific use of the imagination."} (Conan Doyle, 1902), the picture that emerges from the whole analysis is that there are two distinct line forming regions contributing to the observed spectrum at any time: the low and the high velocity component. 

We identify the HVC as the mass outflow produced with the 2008 outburst. Its expansion velocities are likely larger than observed because we are viewing the binary at some intermediate inclination (see below), although the measured radial velocity (up to $\sim-$400 km/s) is augmented by the systemic radial velocity ($\gamma\sim-$120 km/s)\footnote{For $i\sim$45$^o$ and taking into account the systemic radial velocity, the outflow velocity could be as large as $-$400 km/s.}. The extremely peaked, narrow profile that is offset with respect to $\gamma$, of the HVC once in emission, is consistent with sudden, time limited mass loss in a rapidly accelerating, possibly somewhat collimated, wind (for instance, as modeled for the outbursting symbiotic MWC560 by Shore et al., 1994); the peak of the emission line corresponding to where the velocity approaches saturation,  its width resulting from the velocity gradient. The missing "red side" indicates  the presence of dust obscuring the receding part of the outflow in an intermediate inclination system: dust in bipolar ejecta viewed edge-on cannot asymmetrically affect the line profile; conversely, the inclination does not play a role in the observed asymmetry only in case of spherical ejecta. Ground truth about the dust effect comes from the inspection of the sub-mm SiO and CO line profiles in Kaminski et al. (2018, their figure 8 in particular), which are immune to extinction and show the whole emission region with its receding part. 
Finally, part of the emission just redward of about $-$200 km/s, is also blocked by the LVC and argues some geometrical thickness of the latter. 
The HVC shows a ionization gradient with the highest energy transitions in its inner/low velocity layers, and the low energy emission more in its outskirt. This ionization structure is similar to the nested ionized regions in a planetary nebula and is also consistent with a density distribution produced by rapidly accelerating wind of relatively short duration; it also demonstrates the existence of a central ionizing source. 
Initially very opaque, the HVC turns transparent only after the central source has "turned off", so that nothing but forbidden transitions from low ionization potential energy metals can develop. After about one year from the outburst the HVC shows a density of several 10$^6$ cm$^{-3}$. 

The LVC matches circumbinary gas with a substantial column density and which absorbs and is excited in its inner layers by the outburst (i.e. the gas is ionization bounded). Its fluorescence, observed particularly well in the [O~I] and [Ca~II] lines, is similar to that of the SN 1987A ring (Dwek \& Felten 1992). The time required to reach the fluorescence maximum depends on the distance of the gas from the central source, the duration of the central source pulse, and the physical conditions of the gas and its local emissivity. This implies that we cannot disentangle the light delay time (to reach the gas) and the diffusion time (within the gas), but the maximum flux is observed once the light from the whole emitting volume has reached the observer.  This happens in epoch 5 indicating a representative time scale for the combined delay+diffusion time of $\sim$35 days. Without the delay time we cannot precisely determine the distance of the LVC from the central source or when it left the system, but an upper limit for its distance can be  estimated assuming no diffusion time, so that 35/2 light days, i.e. $\sim$3000 AU, is the LVC standoff distance. We know, however, that the diffusion time is not negligible since the [O~I]$\lambda\lambda$6300,6364 absorptions were already present in the first spectrum indicating that the light pulse had already reached the LVC in less than 10 days. Without knowing the delay time it is not possible to accurately compute the LVC density, but the mere presence of the [O~I] emission lines and the $\lambda$5577 in particular, argues for values not much larger than the critical density, $n_e\sim$10$^8$ cm$^{-3}$. This value is consistent with that derived in Section~\ref{sect:pulse} from the recombination timescale of the He~I, implying that the LVC is stationary in the time interval of our observations and further supporting the scenario in which the LVC is distinct from the transient (i.e. central object plus "ejecta") and unaffected by it other than by fluorescence. The large column density (3-5$\times$10$^{23}$ cm$^{-2}$) suggest a substantial mass loss from past times. When combined with the O~I $\lambda$5577 critical density ($\sim$4$\times10^{23}$/10$^8$) the radial extension of the LVC is as large as $\sim$260 AU,  suggesting long formation time scales. Taken together, this is in line with McCollum et al. (2014) analysis of the Spitzer preoutburst data whose color matches those of  "highly evolved stars with large amount of chemically complex mass loss, such as AGB stars". But the LVC is even more complex since it shows, in addition, (see Na~I profile in Fig.\ref{fig:barycenter}) non-saturated narrow absorption components at lower velocity (in the range $-140$,$-100$ km/s) which might arise from mass lost from the Lagrangian point L$_2$ and the spiral structure predicted by theoretical models (e.g. Webbink 1976, Pejcha 2016a, 2016b, 2017,  MacLeod \& Loeb 2020). The fact that they are observed only in the Na~I transitions indicate a significantly lower column density than in the main LVC absorption.

There are no signs of dynamical interaction between the two components. All the observed emission processes are consistent with photon processes and not with collisions: from the absorbed blue ($<$4000 \AA) continuum suggestive of Rayleigh scattering by the HVC, to the HVC ionization structure, to the fluoresclent emission lines in the LVC. This lack of dynamical interaction signatures suggests that the LVC and the HVC are orthogonal to each other. 
The alternate scenario of two distant concentric shells seems less likely, for, otherwise, we should not be able to observe the HVC while the low velocity gas is opaque (which it always is), nor would we see its excited inner layers -- unless it is extremely fragmented and patchy. The line profiles do not indicate this. The LVC, however, is geometrical thick, since it intercepts part of the light from the HVC. 
We, therefore, propose that the HVC is a somewhat bipolar rapidly accelerating wind, and the LVC has a torus-like geometry. Although V1309 Sco does not yet appear convincingly elongated in submm ALMA images (and we suggest this might be due to its distance in addition to its age),  bipolar geometries have been inferred for similar objects from submm observations (Kaminski et al 2018). 

In summary, we have observationally obtained a description of the structure and geometry of V1309 Sco during the 2008 outburst, and inferred approximate densities for the emitting volumes and qualitative dynamics. We also provided evidence for a high energy cutoff of the emission from the central source, but we cannot further constrain its bolometric luminosity or its precise spectral energy distribution (SED). Nonetheless, whatever the energy generation mechanism(s) was and however it varied, the central source certainly turned off on time scales of
months. It emitted in the UV range but spectroscopic observations were not taken in this energy band during V1309 Sco event. This spectral range is especially important during the rise time, since it would have provided a stronger constraint of the SED, the start time of the event, and when the ejecta turned UV optically thick. In light of the column density we inferred for the V1309 Sco circumbinary material, the curtain would have sufficient opacity to completely redistribute
the UV flux to the optical if the source was completely covered. Hence, prompt FUV and NUV observations should be attempted for the next candidate merger since the relative emissivity of the UV and optical flux depends on the source coverage, if independent reddening estimates can be obtained. The same argument goes for the IR and dust redistribution and is independent of the distance. Also, V1309 Sco optical light curve is photometric redistribution by the mass expelled with the outburst and by the circumbinary gas as they recombine. Since the reprocessing material is not spherical, the observed optical light is only a fraction of the central source bolometric emission. 

V1309 Sco, might have been a peculiar event, but it is the only certain noncompact binary star merger we know. It is therefore important to consider all of the above constraints when producing theoretical models of noncompact stars mergers.
Stellar mergers resulting from a common envelope event have been modeled by numerous authors both as general case (see Ivanova et al. 2013 for a comprehensive review; see also Pejcha et 2016a, 2016b, Metzger \& Pejcha 2017, MacLeod \& Loeb 2020 for more recent developments) and  simulated specifically for V1309 Sco (e.g.  Nandez et al. 2014, Pejcha et al. 2017 and references therein). Direct comparison between those models and our results is not straightforward since theoretical codes are based on a number of simplifications that might not apply to V1309 Sco specific case (e.g. the point mass plunging in the extended envelope of a giant star), and seek to reproduce photometric behaviors which might be incomplete and are notoriously ambiguous. Still, the following points are readily noticeable.

1) All the models predict a substantial mass loss during the phases preceding the dynamical merger and resulting in the presence of circumbinary material with a toroidal or disk-like geometry. While we confirm the presence of such a circumbinary material, its extension and formation timescale are not quite in agreement with the theoretical predictions.  Our circumbinary material is well detached from the central binary as shown by the time delay between the appearance of the blue shifted LVC absorption and its emission counterpart. The simulated medium is much more compact, relatively close or even  connected to the binary with scale of tens of giant-star radii or orbital separations, and piled up on time scales of years at the very most. Observationally, for a standoff distance of $\leq$3000 AU and a drift velocity of $\sim$60 km/s, the gas left the system hundreds of years before the outburst.  The independent estimate of the gas extension along the orbital plane of $\simeq$260 AU, also argue for long formation time scales and is orders of magnitude larger than that predicted by the simulations (e.g. compare with the $\sim$60 binary separation, $a$, or stellar donor radii, $R_1$, in Figures 9 of Pejcha et al. 2017 and 5 of MacLeod \& Loeb 2020). The observed column density is also order of magnitude smaller than predicted by Pejcha et al. (2017) simulations.

2) In the attempt to reproduce the observed light curves of V1309 Sco and alike transients, Metzger \& Pejcha (2017) proposed shocks between the slowly moving outflow from L$_2$ and the fast ejecta launched during the dynamical phase of the merger. In particular, second light curve peaks or plateau result depending on the shock energetic capable (or less) to keep hydrogen ionized. Our non detection of any signature of dynamical interaction between, and within, the HVC and the LVC in the time span of our observations (392 days) dismiss the proposed scenario for V1309 Sco. It will be important to monitor through sufficient high resolution spectroscopy future candidate stellar mergers to collect kinematic and dynamical information of the expanding gas(es) during their major photometric phases. Here, we recall the light echo effect produced by circumstellar dust that is off the line of sight and that might have same, bluer or redder color depending on the differential extinction between the scattering dust and the transients. This is fairly well known for supernovae (e.g. Patat 2005) and has also been modeled for classical novae (Shore 2013). Ultimately only (spectro)polarimetry and line profile evolution studies can serve as a diagnostic for the mechanisms behind a given light curve. Imagery and bolometry alone are insufficient. 

We conclude by noting the importance of considering the observed spectral signatures when defining the parameter space to be explored in simulations. Models and simulations should predict  spectroscopic observables.  
Ideally, detailed treatment of the relevant radiative processes should be included in the common envelope studies to predict the observed spectrum and its time evolution. 

\begin{acknowledgements}
      Based on observations collected at the European Organisation for Astronomical Research in the Southern Hemisphere under ESO programmes 281.D-5055(A) and 60.A-9445(A). 
      
      This research has made use of NASA’s Astrophysics Data System Bibliographic Services. 
      
      This work also made extensive use of the {\it NIST atomic data base and online tool} and the van Hoof' {\it Atomic line list v3.00b4} online tool. It also used the {\it Partial Grotrian diagram of astrophysical interest} by Moore \& Merill (1968, NSRDS - NBS 23).
      We thank the referee, Tomasz Kaminski, for comments and suggestions that improved the paper. EM also thanks the  Lodovico Coccato and Christian Hummel from ESO USD and DMO for their help with {\it molecfit}.
\end{acknowledgements}

% WARNING
%-------------------------------------------------------------------
% Please note that we have included the references to the file aa.dem in
% order to compile it, but we ask you to:
%
% - use BibTeX with the regular commands:
%   \bibliographystyle{aa} % style aa.bst
%   \bibliography{Yourfile} % your references Yourfile.bib
%
% - join the .bib files when you upload your source files
%-------------------------------------------------------------------

\end{document}